# Disorder-mediated linear and nonlinear magnetotransport in the charge-density-wave material Ta$_2$NiSe$_7$

Xiaodong Sun[1,2], Jiabin Qiao[1,2,*], Yuanzhe Li[1,2], Wanli He[1,2], Jiali Chen[1,2],
Jinjin Liu[1,2], Yuxiang Chen[1,2], Yuchen Ma[1,2], Meiling Jin[1,2], Jianlin Luo[3,4,5],
Jie Chen[1,2,*], Wei Wu[3,*], Zhiwei Wang[1,2,*], Wei Jiang[1,2,*], Xiang Li[1,2,*], Yugui Yao[1,2]

[1]*Centre for Quantum Physics, Key Laboratory of Advanced Optoelectronic Quantum Architecture and Measurement (MOE), School of Physics, Beijing Institute of Technology, Beijing, 100081, China.*
[2]*Beijing Key Lab of Nanophotonics & Ultrafine Optoelectronic Systems, School of Physics, Beijing Institute of Technology, Beijing, 100081, China.*
[3]*Beijing National Laboratory for Condensed Matter Physics, and Institute of Physics, Chinese Academy of Sciences, Beijing 100190, China*
[4]*School of Physical Sciences, University of Chinese Academy of Sciences, Beijing 100190, China*
[5]*Songshan Lake Materials Laboratory, Dongguan 523808, China*

\* Email:
jiabinqiao@bit.edu.cn
jiechen.phy@bit.edu.cn
welyman@iphy.ac.cn
zhiweiwang@bit.edu.cn
wjiang@bit.edu.cn
xiangli@bit.edu.cn

## Abstract

We report disorder-mediated first-order linear and higher-order nonlinear (magneto-)transport of Ta$_2$NiSe$_7$ (TNS) in the charge-density-wave (CDW) regime. CDW transition temperature ($T_{\text{CDW}}$) and carrier density are proportional and inversely proportional to residual resistance ratio of samples, respectively. Such relation helps to understand the unique CDW order therein. High-$T_{\text{CDW}}$ TNS exhibits negative first-harmonic magnetoresistance ($\text{MR}^{1\omega}$) under a magnetic field ($B$) parallel to the direction of alternating current ($I^\omega$), which may arise from the anomalous velocity induced by the Berry curvature of three-dimensional topological bands near the Fermi level. As $T_{\text{CDW}}$ drops, a positive-to-negative $\text{MR}^{1\omega}$ transition is observed with decreasing perpendicular $B$, which is likely due to the contribution of Zeeman effect on current pathways in the disordered system. Moreover, interestingly, the second-harmonic nonlinear signals are suppressed, while the third-harmonic signals are significant and sensitive to both $B$ and $T_{\text{CDW}}$. Such observations, together with scaling analysis, suggest the quantum geometry quadrupole at play and the modulation of disorder on third-order nonlinearity. Our results pave an avenue for tailoring distinct-order magnetoresistive phases in disordered topological materials.

It is straightforward to detect a quadratic positive magnetoresistance (pMR) in most metals and semiconductors due to magnetic field ($B$)-induced localization of electrons in cyclotron orbits [1,2]. However, the magnitude of such pMR is always weak and hampers practical applications such as magnetic field sensing [3]. By contrast, diverse topological materials showcase exotic MR phenomena thanks to unique topological orders [4-11], which helps to unveil fundamental physics and implements various functions for applications. Apart from nontrivial band topology, disorder induced by defects/impurities acts as a tuning knob for modulating the MR behaviors in such materials [12-14]. Control over the disorder is of great importance for elucidating and tailoring distinct magnetoresistive phases. Most recently, research interests have shifted to higher-order nonlinear transport in various materials [15-23], which is intimately linked to quantum geometry effects [24-27]. However, it remains largely unexplored how disorder affects such nonlinear magnetotransport, especially the third-order nonlinearity.

Here, we present systematic magnetotransport study of ternary chalcogenides Ta$_2$NiSe$_7$ (TNS) featuring a variable CDW transition temperature ($T_{\mathrm{CDW}}$). We unveil the relations among $T_{\mathrm{CDW}}$, carrier density ($n_s$), and residual resistance ratio (RRR) of samples, which help to understand the unique CDW order. Bulk TNS exhibits $T_{\mathrm{CDW}}$-specific anisotropic first-harmonic MR ($\mathrm{MR}^{1\omega}$) behaviors, where the negative $\mathrm{MR}^{1\omega}$ ($\mathrm{nMR}^{1\omega}$) is governed by different mechanisms. Moreover, the second-harmonic nonlinear signals are suppressed while the third- harmonic signals are prominent and of significant $B$ dependence, suggesting the quantum geometry quadrupole at play under inversion symmetry accompanied with other extrinsic contributions. Strikingly, the $\mathrm{MR}^{3\omega}$ exhibits non-monotonic $B$ dependence and its amplitude varies one to three order(s) of magnitude with decreasing $T_{\mathrm{CDW}}$, indicating the modulation of disorder on higher-order moments of quantum geometry and resultant nonlinear transport.

Transition metal ternary chalcogenides, Ta$_2$NiSe$_7$ (TNS), have long garnered much attention due to complex crystalline structure, controversial topological character, and peculiar CDW phase [28-40]. Overall, it is a layered material in a centered monoclinic space group (No. 12, *C*2/*m*), despite a staggered van der Waals gap normal to the *b-c* plane, as shown in Fig. 1(a). Looking over each layer, it can be viewed as a quasi-1D system consisting of three types of atomic chains: one comes from the Ni atoms (marked by blue balls) in distorted octahedra (d-OCT) form and the others are double chains formed by Ta1 (red balls) and Ta2 (purple balls) atoms in bicapped trigonal prismatic (BTP) and octahedral (OCT) coordination, respectively. In such "2+1D" crystal, the predicted band topology is currently under dispute in either the CDW ($T < T_{\text{CDW}}$) or the normal-state ($T > T_{\text{CDW}}$) regime [35-37,57,58], and also incompatible with recent angle-resolved photoemission spectroscopy (ARPES) results [38]. More prominently, bulk TNS hosts two distinct CDW wavevectors, $\mathbf{q} = (0,0.483,0)$ and **2q**, at which lattice distortion is manifested as transverse and longitudinal displacement of Ni/Se and Ta2 atoms, respectively [30-33,38]. However, the understanding of such unique CDW order and its origin remains elusive [33,38,39].

In our experiments, tens of TNS flakes are exfoliated from different batches of single crystals for transport measurements (Supplemental Material, Methods, Table S1, and Figs. S1-S4 [41]). The distinct-harmonic longitudinal ($R_{xx}^{1,2,3\omega}$) and transverse ($R_{yx}^{1,2,3\omega}$) resistance are obtained by measuring the corresponding voltages $V_{xx}^{1,2,3\omega}$ and $V_{yx}^{1,2,3\omega}$ driven by an alternating current ($I^{\omega}$) with the same frequency $\omega$. Figure 1(b) shows $T$ dependence of normalized first-harmonic resistance $R_{xx}^* = R_{xx}^{1\omega}(T)/R_{xx}^{1\omega}(300\text{ K})$ in a representative sample (S9). Upon cooling, the monotonically decreasing trend confirms its metallic nature, in agreement with the results reported in previous works [34-39]. Notably, a kink near 60.8 K signals the onset of CDW transition. We take the derivative of normalized resistance ($dR_{xx}^*/dT$) and define $T_{\text{CDW}}$ from the temperature point at which $dR_{xx}^*/dT$ reaches its dip. Figure 1(c) shows $dR_{xx}^*/dT$ of seven samples as a function of $T$, where the dip moves towards the lower $T$. We summarize the extracted $T_{\text{CDW}}$ and $n_s$ (Fig. S5 in Supplemental Material [41]) versus RRR= $R_{xx}^{1\omega}(300\text{ K})/R_{xx}^{1\omega}(2\text{ K})$, as plotted in Fig. 1(d).

Notably, $T_{\text{CDW}}$ and $n_s$ are proportional and inversely proportional to RRR, respectively, indicating that the disorder boosts *p*-type carrier doping and modulates the CDW transition in TNS. Such doping effect is unlikely associated with the CDW gap that typically varies with $T_{\text{CDW}}$, since the gap cannot be detected via ARPES measurement [38] and previous transport results show that the carrier density keeps constant around $T_{\text{CDW}}$ [34]. Such doping effect is likely due to the presence of metal-atom vacancy defects, which always act as hole donors [59]. The modulation of $T_{\text{CDW}}$ is similar to that observed in Nb-doped TNS, $(Ta_{1-x}Nb_x)_2NiSe_7$ [33], where the tuning mechanism is tied to the **2q** ordering at the Ta2 atoms instead of the primary **q** ordering at the Ni/Se sites. Given these facts, we infer that such linear relations originate from the presence of Ta2 (vacancy) defects. As RRR drops, the Ta2 vacancies accumulate and give rise to increasing *p*-type carrier density. Meanwhile, they pose a marked impact on the **2q** ordering, resulting in the decrease of $T_{\text{CDW}}$. This speculation is supported by our X-ray energy dispersive spectroscopy (EDS) results where the Ta-atom concentration in low- (high-) $T_{\text{CDW}}$ samples is statistically less (more) than 20 % (Fig. S3 in Supplemental Material [41]). All results highlight the crucial role of the **2q** ordering in the formation of CDW in TNS.

First-harmonic magnetotransport is investigated in high-$T_{\text{CDW}}$ samples. Figure 2(a,b) shows the first-harmonic MR, $\text{MR}^{1\omega} = [R_{xx}^{1\omega}(B) - R_{xx}^{1\omega}(0)]/R_{xx}^{1\omega}(0) \times 100\%$, of S1 ($T_{\text{CDW}}$ = 67 K, RRR = 5.3) as a function of $B$ applied with an out-of-plane polar angle $\theta = 90°$ and $\theta = 0°$ from the direction of alternating current ($I^\omega$) along the *b* axis of the crystal at various $T$, respectively. Clearly, the quadratic positive $\text{MR}^{1\omega}$($\text{pMR}^{1\omega}$) is observed under either $B_\perp^{op}$ (i.e., $B \perp b$) or $B_\parallel$ (i.e., $B // b$) at $T \geq 8$ K. However, the $\text{nMR}^{1\omega}$ occurs in a wide range of $B_\parallel$ at low $T$. Among various mechanisms for such $\text{MR}^{1\omega}$ behaviors, the anisotropic behavior first rules out the Zeeman effect on percolating current pathways in the disordered system [13], the Kondo effect [49,50] or the 3D weak localization (WL) [51] as the possible origin, since the first two produce isotropic $\text{nMR}^{1\omega}$ while the last gives rise to the dominance (suppression) of $\text{nMR}^{1\omega}$ under $B_\perp^{op}$ ($B_\parallel$), all in stark contrast to our observations (Note 1 and Table S2 in Supplemental Material [41]). Moreover, possible external artifacts such as current jetting are carefully excluded [52] (Fig. S7 in Supplemental Material [41]). Indeed, such $\text{MR}^{1\omega}$ behaviors are strongly reminiscent of those observed in topological semimetals and insulators due to chiral anomaly [6-8] and Berry curvature effect [53,60], respectively.

To gain more insight into the band topology in the CDW regime, we perform full electronic simulations based on first-principles density functional theory (DFT) (Methods, Note 2 and Figs. S9, S10 in Supplemental Material [41]). Our theoretical calculations explicitly demonstrate that bulk TNS is a 3D topological insulator (TI) with Z4=3 while considering spin-orbit coupling (SOC) [55], consistent with that proposed in the normal state [58]. Interestingly, we find that the magnitude and tendency of $\mathrm{nMR}^{1\omega}$ in S1 exactly coincide with that in the strong TI $Bi_2Se_3$ [60] (Fig. S8 in Supplemental Material [41]). The overlapped data are in quantitative agreement with the theoretical calculation taking into account the correction to magnetoconductivity from the Berry curvature [53]. Therefore, we attribute such $\mathrm{nMR}^{1\omega}$ to the Berry-curvature-induced anomalous velocity and resulting enhancement of conductivity under $B_\parallel$. To further elaborate on this point, angular dependence of $\mathrm{MR}^{1\omega}$ of S1 and S4 ($T_{\mathrm{CDW}}$ = 64 K, RRR = 4.38) is further investigated by rotating the sample out of [Fig. 2(c)] and in [Fig. 2(d)] the *b-c* plane with respect to applied *B*, respectively. Notably, the total $\mathrm{MR}^{1\omega}$ at any angle can be quantitatively treated as a superposition of $\mathrm{nMR}^{1\omega}$ and $\mathrm{pMR}^{1\omega}$ components that depends exclusively on the parallel and perpendicular field components, respectively (Note 1 and Fig. S6 in Supplemental Material [41]). This excellent agreement with no free parameters provides robust, quantitative evidence that the $\mathrm{nMR}^{1\omega}$ is indeed a function only of $B_\parallel$. We remark that, despite the presence of nontrivial topology, the topological bands traverse the Fermi level, thus resulting in a metallic behavior.

We then turn to the first-harmonic magnetotransport in low-$T_{\mathrm{CDW}}$ samples. Figure 3(a,b) shows $B_\perp^{op}$ and $B_\parallel$ dependence of $\mathrm{MR}_{xx}^{1\omega}$ in S17 ($T_{\mathrm{CDW}}$ = 50 K, RRR = 2.92), respectively. Similar $\mathrm{nMR}^{1\omega}$ behavior is observed under $B_\parallel$ [Fig. 3(b)], suggesting the Berry curvature contribution still at play. Whereas, an unexpected $\mathrm{nMR}^{1\omega}$ occurs in a small range of $B_\perp^{op}$ (< 5 T) at $T < 8$ K [Fig. 3(a)]. Such $\mathrm{nMR}^{1\omega}$ behavior under $B_\perp^{op}$ is more likely due to the Zeeman effect on percolating current pathways in such disordered bulk (referred to as disorder contribution to $\mathrm{nMR}^{1\omega}$ below) [13], rather than 3D weak localization (WL) effect [51] (Fig. S11 in Supplemental Material [41]). To elucidate the $MR^{1\omega}$ behaviors in bulk TNS with different $T_{CDW}$, a circuit model is employed for phenomenological simulation, as illustrated in Fig. 3(c). The equivalent circuit mimicking realistic system is comprised of two resistors ($R_p$ and $R_n$) in series, where the total resistance is described as [61]:

$$R(B) = R_p(B) + R_n(B) = R_0[\gamma(1 + \alpha B^2) + (1 - \gamma)/(1 + \beta B^2)], \quad (1)$$

where $R_0$ is the measured resistance at zero field, $\alpha$, $\beta$ and $\gamma$ are fitting parameters. The first and second terms denote the positive and negative contributions to the total MR with a ratio of $\gamma$ and $1 - \gamma$, respectively.

The $MR^{1\omega}$ behaviors in both high- and low-$T_{CDW}$ samples are well captured by using the phenomenological model, as shown in Fig. 3(d-f) (Figs. S12-S17 in Supplemental Material [41]). Figure 3(d,e) plots $T$ dependence of the derived fitting parameters for the high-$T_{CDW}$ sample S3 ($T_{CDW}$ = 65 K, RRR = 4.29) and the low-$T_{CDW}$ one S17 under $B_\perp^{op}$ and $B_\parallel$, respectively. At low $T$ ($< 10$ K), the $nMR^{1\omega}$ component in S3 accounts for more than 80 % under $B_\parallel$, while it is negligible under $B_\perp^{op}$ [Fig. 3(d)], which is an indication of the $nMR^{1\omega}$ caused by the Berry curvature effect. As $T_{CDW}$ (or RRR) drops, disorder-driven $nMR^{1\omega}$ component contributes to the first-order magnetotransport under $B_\perp^{op}$ and reinforces the dominance of $nMR^{1\omega}$ under $B_\parallel$ [Fig. 3(e)]. We further determine angle ($\theta$) dependence of the ratio of pMR to nMR at low $T$. In the high-$T_{CDW}$ sample S1, the Berry-curvature-induced $nMR^{1\omega}$ significantly relies on the $B$ component parallel to $I^\omega$ such that its ratio plummets and rapidly vanishes with increasing $\theta$ [Fig. 3(f), upper]. When the disorder contribution is involved, the $nMR^{1\omega}$ in the low-$T_{CDW}$ sample S10 ($T_{CDW}$ = 55 K, RRR = 2.44) survives at much larger $\theta$ since the disorder contribution behaves in an isotropic fashion [Fig. 3(f), lower].

Apart from the tunable first-order magnetotransport, bulk TNS exhibits distinct higher-order nonlinear (magneto-)transport behaviors in the CDW regime. After carefully excluding possible current heating effect (Fig. S18 in Supplemental Material [41]), we systematically investigate transport behaviors in distinct harmonics by injecting moderate $I^\omega$ into different samples. Figure 4(a) plots distinct-harmonic longitudinal voltages $V_{xx}^{1,2,3\omega}$ in S8 ($T_{CDW}$ = 62 K, RRR = 5.25) as a function of $I^\omega$ at $T = 2$ K and $B = 0$ T. Obviously, the trace of $V_{xx}^{1\omega}$ shows an excellent linearity, conforming to Ohm's law under $I^\omega \leq 1000$ μA. By contrast, $V_{xx}^{2\omega}$ and $V_{xx}^{3\omega}$ scale quadratically and cubically with $I^\omega$, which is a direct indication that the second- and third-order quantum nonlinearity occurs in bulk TNS. Note that the second-harmonic nonlinear signals are around twice lower than the third-harmonic ones. We further investigate $B_\perp^{op}$ dependence of $V_{xx}^{1,2,3\omega}$ in distinct-$T_{\text{CDW}}$ samples, as shown in Fig. 4(b,c). Remarkably, $V_{xx}^{2\omega}$ has a relatively small magnitude and varies slightly with $B_\perp^{op}$, while $V_{xx}^{3\omega}$ is prominent and displays drastically different magnetic field dependence.

We here discuss the origin of such higher-order magnetotransport. The third- (second-) harmonic nonlinear signals are prominent (suppressed), indicating the dominance of third-order nonlinearity over second-order harmonics in bulk TNS where inversion symmetry is substantially retained. The scaling relation of third-order nonlinearity under time-reversal symmetry can be described as [21,22] (Note 3 in Supplemental Material [41]):

$$V_{xx}^{3\omega}/(V_{xx}^{1\omega})^3 = p\sigma_{xx}^2 + q, \qquad (2)$$

where $\sigma_{xx}$ is first-harmonic longitudinal conductivity, the first and second terms on the right-hand side denote the Drude-like tensor and quantum metric quadrupole (QMQ) contributions, respectively. The fitting of Eq. (2) to the experimental data in the high- and low-$T_{\text{CDW}}$ samples, S2 ($T_{\text{CDW}}$ = 66 K, RRR = 3.7) and S19 (with no CDW transition), is plotted in Fig. 4(d,e), respectively. Clearly, the low-$T$ data (corresponding to the higher $\sigma_{xx}$) agree well with such scaling law, revealing that both Drude-like tensor and QMQ contribute to the third-order nonlinearity with comparable magnitude. Interestingly, both the contributions decrease with $T_{\text{CDW}}$. Notably, the obtained QMQ contribution ($q$) in S2 at low $T$ reaches around 4700 V$^{-2}$, which is two orders of magnitude larger than that in antiferromagnetic TI bulk MnBi$_2$Te$_4$ [21] and few-layer WTe$_2$ [22]. We remark that the data deviate from the linear fit at higher $T$ [or lower $\sigma$, see Fig. 4(d,e)], which indicates that other $\tau$-dependent conductivity mechanisms than the Drude-like and QMQ contributions play a role in the third-order nonlinear transport. Such deviation is likely due to the enhancement of phonon skew scattering process and suppression of the $\tau$ term (here is QMQ) with increasing $T$. Similar behavior has been observed in few-layer WTe$_2$ [22].

Moreover, the measured $V_{xx}^{3\omega}$ (or $\text{MR}^{3\omega}$) is an even function of $B_\perp^{op}$ [Fig. 4(b,c)] while the $V_{yx}^{3\omega}$ is odd in $B_\perp^{op}$ (Fig. S19 in Supplemental Material [41]). In the third-order nonlinearity, the quantum metric (Berry curvature) quadrupole contributes exclusively to the $V_{xx}^{3\omega}$ ($V_{yx}^{3\omega}$) with even- (odd-)symmetry with respect to the directions of applied $B_\perp^{op}$. Whereas, the contribution from Drude-like tensor (skew scattering) is insensitive to $B_\perp^{op}$ [21]. Therefore, the $B$ dependence of third-order nonlinearity is mainly governed by intrinsic QMQ with a constant background formed by extrinsic contributions. Nevertheless, the origin of non-monotonic $B$-dependent $V_{xx}^{3\omega}$ (or $\text{MR}^{3\omega}$) remains unknown and warrants further theoretical and experimental investigations.

To better compare the higher-order magnetotransport behaviors with the first-order ones, we define the higher-harmonic MR as $\text{MR}^{2,3\omega} = [V_{xx}^{2,3\omega}(B) - V_{xx}^{2,3\omega}(0)]/V_{xx}^{2,3\omega}(0) \times 100\%$. Figure 4(f) plots $B$ dependence of $\text{MR}^{1,2,3\omega}$ in S8 and S15 at $T = 2$ K. Obviously, the amplitude of $\text{MR}^{3\omega}$ is much larger than that of $\text{MR}^{1\omega}$ or $\text{MR}^{2\omega}$, and its polarity is tuned by both $B$ and $T_{\text{CDW}}$. Furthermore, we summarize the maximum of $\text{MR}^{1\omega}$ and $\text{MR}^{3\omega}$ under various $I^\omega$ within 10 T in multiple samples as a function of $T_{\text{CDW}}$, as plotted in Fig. 4(g). Despite a slight fluctuation of $\text{MR}^{1\omega}$, the $\text{MR}^{3\omega}$ expands exponentially with decreasing $T_{\text{CDW}}$, indicating a significant effect of disorder on third-order nonlinear magnetotransport.

In summary, we carry out first-order linear and higher-order nonlinear (magneto-)transport study in bulk TNS. The RRR dependence of $T_{\text{CDW}}$ and $n_s$ suggests a crucial role of the **2q** ordering located at Ta2 sites in the formation of CDW. The $\text{nMR}^{1\omega}$ in high-$T_{\text{CDW}}$ samples relies exclusively on $B//I^\omega$, whose origin is tied to the Berry curvature effect. As $T_{\text{CDW}}$ drops, a growing degree of disorder brings into play an additional contribution from the Zeeman effect on percolating current pathways. The whole $\text{MR}^{1\omega}$ behaviors can be well captured by a phenomenological two-resistor model. Moreover, the suppressed second-order nonlinear signals enable the third-order ones—mainly contributed by both intrinsic QMQ and extrinsic Drude-like tensor—standing out as the leading nonlinear longitudinal response under inversion symmetry. The $\text{MR}^{3\omega}$ is giant and its magnitude is sensitive to $T_{\text{CDW}}$. Our work not only advances the understanding of disorder-mediated linear and nonlinear (magneto-)transport as well as associated quantum geometric properties, but also opens a feasible route for achieving field-sensitive device applications based on higher-order MR engineering.

**Figures**

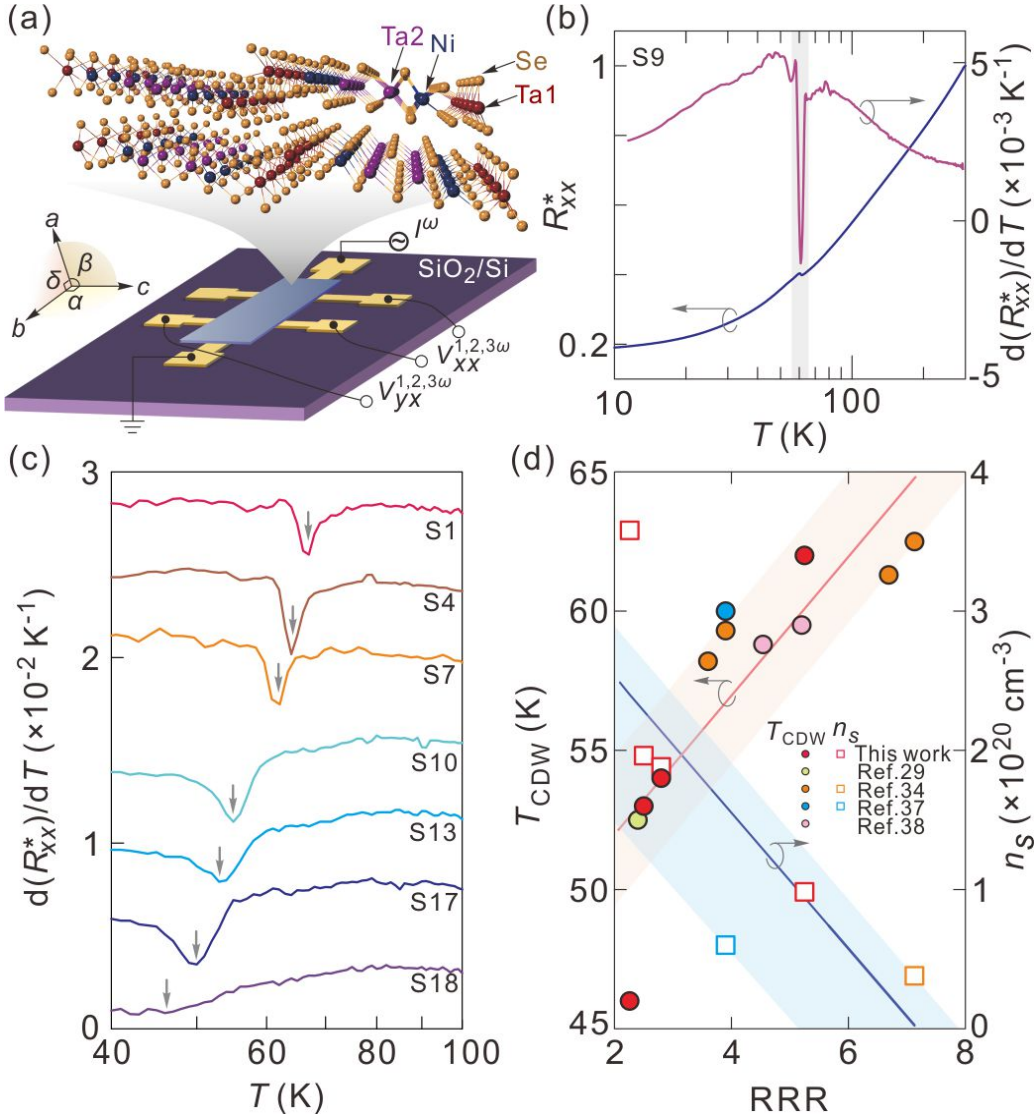

**Figure 1. First-harmonic transport properties of Ta$_2$NiSe$_7$. (a)** Crystal structure and device configuration of Ta$_2$NiSe$_7$ (TNS) where a quasi-one-dimensional chain-like structure is along the *b* axis, $\alpha = \delta = 90°$ and $\beta = 108.8°$. The alternating current ($I^\omega$) is applied along the *b* axis, and the first-, second- and third-harmonic longitudinal ($V_{xx}^{1,2,3\omega}$) and transverse ($V_{yx}^{1,2,3\omega}$) voltages signals are simultaneously detected. **(b)** $T$ dependence of normalized longitudinal resistance $R_{xx}^*$ and its derivative $dR_{xx}^*/dT$ of the sample S9. **(c)** $dR_{xx}^*/dT$ of seven samples versus $T$. Gray arrows mark the CDW transition temperature $T_{\text{CDW}}$. **(d)** $T_{\text{CDW}}$ and carrier density $n_s$ (taken at 2 K) versus RRR. Filled circles and empty squares denote the data of $T_{\text{CDW}}$ and $n_s$ obtained from this work (S8, S11, S13, and S18) and former studies. The red and blue solid lines are linear fits to experimental data, respectively.

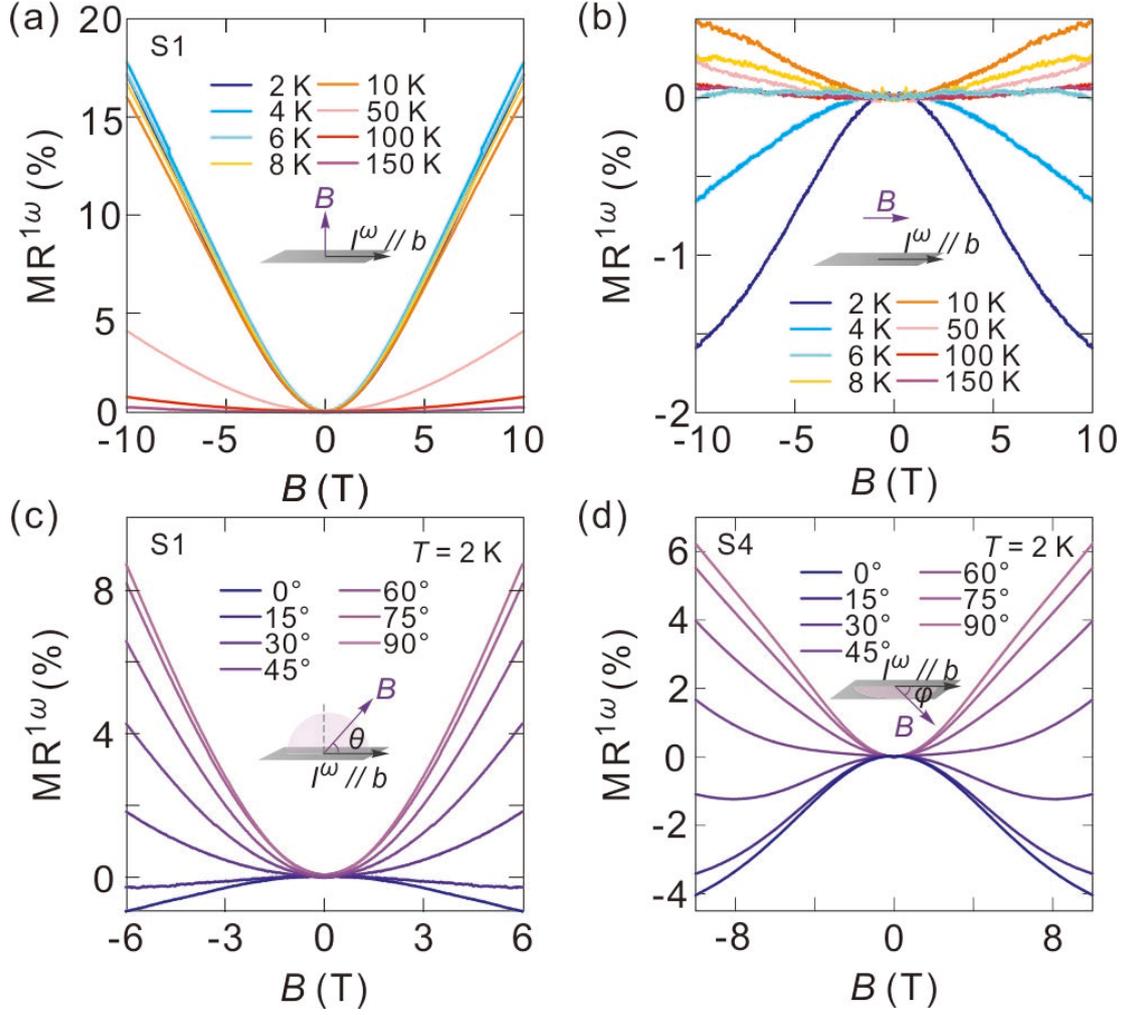

**Figure 2. First-harmonic magnetotransport properties in the high-$T_{CDW}$ TNS flakes.** First-harmonic MR ($MR^{1\omega}$) of S1 versus $B$ applied **(a)** perpendicular to the *b-c* plane (i.e., $B_\perp^{op}$) and **(b)** parallel to the *b* axis (i.e., $B_\parallel$) at a set of $T$. First-harmonic longitudinal MR of S1 **(c)** and S4 **(d)** versus $B$ applied with the rotated angle $\theta$ and $\varphi$ from the direction of $I^\omega$ (or the *b* axis) out of and in the *b-c* plane, respectively. Clearly, the $nMR^{1\omega}$ relies on the $B$ components parallel to $I^\omega$.

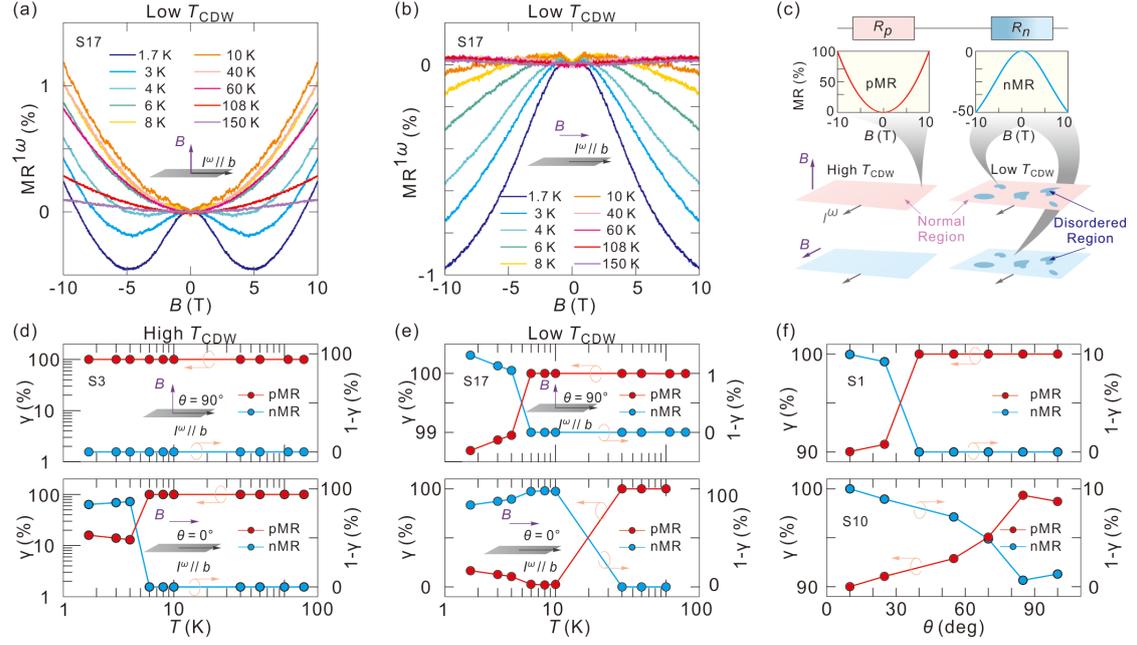

**Figure 3. First-harmonic magnetotransport properties of the low-$T_{CDW}$ TNS flakes and phenomenological simulation using a reduced two-resistor model. (a,b)** First-harmonic MR ($MR^{1\omega}$) of S17 versus $B_\perp^{op}$ and $B_\parallel$ at a set of $T$, respectively. **(c)** Schematic of the two-resistor-in-series model. The normal regions in high- or low-$T_{CDW}$ samples behave as a resistor ($R_p$) contributing to the pMR under $B_\perp^{op}$ (painted in pink) while acting as $R_n$ leading to the nMR under $B_\parallel$ (light blue) thanks to Berry curvature effect. The disordered regions (dark blue) in low-$T_{CDW}$ samples always act as $R_n$ contributing to the nMR. The pMR and nMR are illustrated using $MR(B) = 1 + \alpha B^2$ and $MR(B) = 1/(1 + \beta B^2)$, respectively. For simulation, $\alpha = \beta = 0.01$. **(d,e)** $T$ dependence of the derived ratio for pMR ($\gamma$) and the nMR ($1 - \gamma$) from the reduced model under $B_\perp^{op}$ and $B_\parallel$ in S3 and S17, respectively. **(f)** $\theta$ dependence of the derived ratio for pMR ($\gamma$) and the nMR ($1 - \gamma$) in S1 and S10.

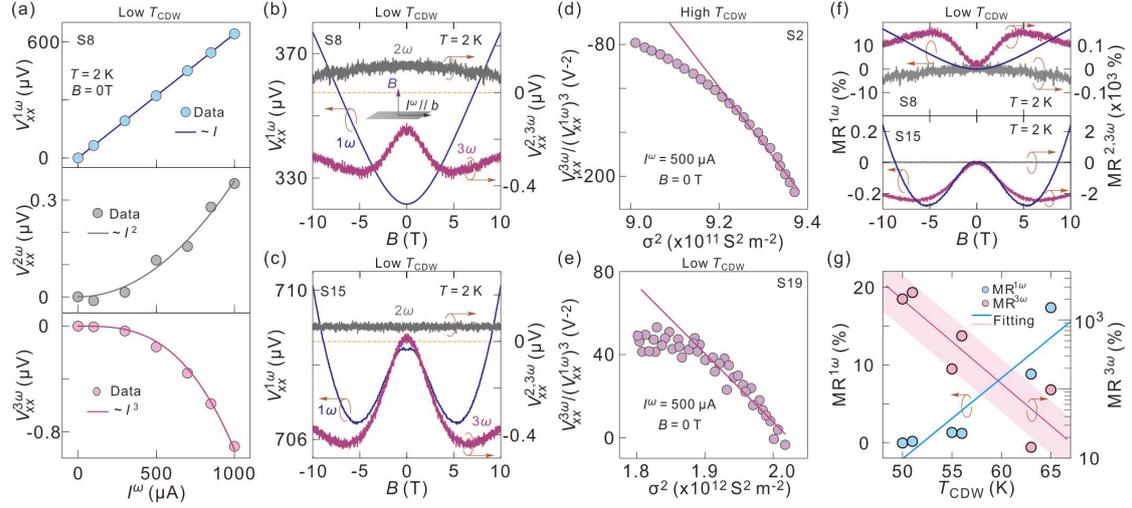

**Figure 4. Higher-harmonic magnetotransport properties in TNS flakes. (a)** Distinct-harmonic *I-V* curves of S8 in the longitudinal direction under zero field at $T = 2$ K. The solid curves indicate the linear, quadratic and cubic fits to experimental data, respectively. **(b,c)** $B$ dependence of $V_{xx}^{1,2,3\omega}$ (marked by blue, gray, purple curves, respectively) in S8 and S15 with $I^\omega = 500$ μA and 70 μA at $T = 2$ K, respectively. **(d,e)** $V_{xx}^{3\omega}/(V_{xx}^{1\omega})^3$ versus $\sigma^2$ in S2 and S19, respectively. The circles represent experimental data. The solid lines are linear fits to the data from 4 K to **(d)** 6.6 K and **(e)** 11 K, respectively. The fitting parameters are: **(d)** $p = -5.21 \times 10^{-9}$ V$^{-2}$Ω$^2$m$^2$, $q = 4672$ V$^{-2}$; **(e)** $p = -3.40 \times 10^{-10}$ V$^{-2}$Ω$^2$m$^2$, $q = 688$ V$^{-2}$. **(f)** $B$ dependence of MR$^{1,2,3\omega}$ (marked by blue, gray, purple curves, respectively) in S8 and S15 at $T = 2$ K. **(g)** The maximum of MR$^{1\omega}$ (marked by purple circles) and MR$^{3\omega}$ (blue circles) data under various $I^\omega$ within ±10 T at $T = 2$ K versus $T_{\text{CDW}}$. The purple and blue lines indicate the exponential ( MR$^{3\omega} \propto e^{-T_{\text{CDW}}/T_0}$, with $T_0 \sim 3.6$ K) and the linear fits, respectively.

**Acknowledgements**

This work is financially supported by the National Key R&D Program of China (Grants Nos. 2023YFA1406002, 2020YFA0308800, 2022YFA1403400, 2022YFA1602802, 2022YFA1602800, 2021YFA1401800), the National Natural Science Foundation (NNSF) of China (Grants Nos. 12321004, 12204045, 12204037, 12134018, 11921004), the Beijing Natural Science Foundation (Grant No. Z210006), and the Beijing National Laboratory for Condensed Matter Physics (Grant No. 2023BNLCMPKF007). J.Q. acknowledges the support by the Beijing Institute of Technology Research Fund Program for Young Scholars. M.J. acknowledges the support by Beijing Institute of Technology Laboratory Research Project (Grant No. 2023BITSYB07).
**Data availability**

The data that support the findings of this article are openly available [62].

calculations and transport data of TNS, which includes Refs. [42-56].